\documentclass[pdflatex,sn-mathphys]{adj_jnl}
\usepackage{graphicx}
\usepackage[inline]{enumitem}

\jyear{2022}%

\begin{document}

\title[Overcoming the disconnect between energy system and climate modelling]{Overcoming the disconnect between energy system and climate modelling}

\affil{For the latest version of the manuscript we refer to \url{https://doi.org/10.1016/j.joule.2022.05.010} }
\affil{$\copyright$ 2022. This manuscript version is made available under the CC-BY-NC-ND 4.0 license \url{https://creativecommons.org/licenses/by-nc-nd/4.0/}}

\author[1]{\fnm{Michael T.} \sur{Craig}}
\equalcont{These authors contributed equally to this work.}
\affil[1]{School for Environment and Sustainability, University of Michigan, USA}

\author*[2,3]{\fnm{Jan} \sur{Wohland}}\email{jan.wohland@env.ethz.ch}
\equalcont{These authors contributed equally to this work.}
\affil[2]{Institute for Environmental Decisions, ETH Zurich, Switzerland}
\affil*[3]{Climate Service Center Germany (GERICS), Helmholtz-Zentrum Hereon, Germany}

\author[4,5,6]{\fnm{Laurens P.} \sur{Stoop}}
\equalcont{These authors contributed equally to this work.}
\affil[4]{Information and Computing Science, and Copernicus Institute of Sustainable Development, Utrecht University, The Netherlands}
\affil[5]{TenneT TSO B.V., Arnhem, The Netherlands}
\affil[6]{Royal Netherlands Meteorological Institute, The Netherlands}

\author[7,27]{Alexander Kies}
\affil[7]{Frankfurt Institute for Advanced Studies, Germany}
\affil[27]{Department of Electrical and Computer Engineering, Aarhus University, Denmark}

\author[8]{Bryn Pickering}
\affil[8]{ETH Z\"urich, Switzerland}

\author[9,10]{Hannah C. Bloomfield}
\affil[9]{School of Geographical Sciences, University of Bristol, UK}
\affil[10]{Department of Meteorology, University of Reading, UK}

\author[11]{Jethro Browell\textsuperscript{11}}
\affil[11]{University of Glasgow, UK}

\author[12]{Matteo De Felice}
\affil[12]{European Commission, Joint Research Centre, The Netherlands}

\author[13,14]{Chris J.Dent}
\affil[13]{School of Mathematics, University of Edinburg, UK}
\affil[14]{The Alan Turing Institute, UK}

\author[15]{Adrien Deroubaix}
\affil[15]{Max Planck Institute, Germany}

\author[16]{Felix Frischmuth}
\affil[16]{Fraunhofer Institute for Energy Economics and Energy System Technology, Germany}

\author[10]{Paula	L.M. Gonzalez}

\author[17]{Aleksander Grochowicz}
\affil[17]{Department of Mathematics, University of Oslo,	Norway}

\author[18]{Katharina Gruber}
\affil[18]{University of Natural Resources and Life Sciences, Austria}

\author[16]{Philipp Härtel}

\author[19]{Martin Kittel}
\affil[19]{German Institute for Economic Research, Germany}

\author[20]{Leander Kotzur}
\affil[20]{Institute of Energy and Climate Research, Techno-economic Systems Analysis, Germany}

\author[21]{Inga Labuhn}
\affil[21]{University of Bremen, Institute of Geography, Germany}

\author[]{Julie K. Lundquist\textsuperscript{22,23,24}}
\affil[22]{Dept. of Atmospheric and Oceanic Science, University of Colorado Boulder, USA}
\affil[23]{National Renewable Energy Laboratory, USA}
\affil[24]{Renewable and Sustainable Energy Institute, USA}

\author[25]{Noah Pflugradt}
\affil[25]{Forschungszentrum Jülich, Germany}

\author[6]{Karin van der Wiel}

\author[26]{Marianne Zeyringer}
\affil[26]{Department of Technology Systems, University of Oslo, Norway}

\author[10]{David J. Brayshaw}


\abstract{Energy system models underpin decisions by energy system planners and operators.
Energy system modelling faces a transformation: accounting for changing meteorological conditions imposed by climate change.
To enable that transformation, a community of practice in energy-climate modelling has started to form that aims to better integrate energy system models with weather and climate models.
Here, we evaluate the disconnects between the energy system and climate modelling communities, then lay out a research agenda to bridge those disconnects.
In the near-term, we propose interdisciplinary activities for expediting uptake of future climate data in energy system modelling.
In the long-term, we propose a transdisciplinary approach to enable development of
\begin{enumerate*}[label=(\arabic*)]
\item energy-system-tailored climate datasets for historical and future meteorological conditions and
\item energy system models that can effectively leverage those datasets.
This agenda increases the odds of meeting ambitious climate mitigation goals by systematically capturing and mitigating climate risk in energy sector decision making.
\end{enumerate*}}


\maketitle

\section{Introduction}
Transforming and decarbonising energy systems is necessary to meet ambitious climate change mitigation goals. This monumental task falls on energy system planners and operators around the world, who share the same primary goal: to provide reliable, resilient, affordable, and clean energy to consumers. System transformation will increasingly couple electric power with other energy sectors~\citep{IPCC15C}. Within individual sectors, system transformation will increasingly couple energy supply and demand with weather and climate. Integrated assessment models provide high-level guidance on the decarbonization needed, but poorly capture meteorological variability~\citep{collins2017integrating}. Higher resolution energy models are the key to capturing this variability within a strongly coupled system~\citep{gernaat2021climate}. The resulting complexity of a strongly coupled system requires increasingly ambitious modelling efforts.

To guide their decisions, energy system planners and operators use energy system models of varying types. As wind and solar power have grown, energy system modelling has undergone a transformation to handle short- and/or long-term meteorological variability and uncertainty~\citep{pfenninger2014energy, lund2017simulation, ringkjob2018review}. This transformation has increased our understanding of how current and future energy systems can be reliable and affordable while relying on variable and uncertain power generation from renewables. Further progress in this area holds one of the keys to successful climate change mitigation~\citep{IPCC15C}.

Despite the recent successes in representing the impact of present-day weather and climate, the energy system modelling community faces another transformation: accounting for the non-stationary meteorological conditions imposed by anthropogenic climate change~\citep{craig2018review, cronin2018climate, yalew2020impacts} and long-term climate variability~\citep{zeng2019reversal, wohland2021mitigating}. An emerging body of literature is highlighting diverse threats that future weather might pose to reliable, resilient, affordable, and clean energy provision. Shifting meteorological and hydrological conditions can affect energy supply, e.g. from renewable~\citep{voisin2016vulnerability, tobin2016climate, reyers2016future, wohland2017more, grams2017balancing, cronin2018climate, karnauskas2018southward, tobin2018vulnerabilities, jerez2019future, vanruijven2019amplification, bartok2019climate, pryor2020climate, hou2021assessing, bloomfield2021quantifying, bloomfield2021importance, gernaat2021climate} and thermal power generation~\citep{vanvliet2012vulnerability, loew2016marginal, miara2017climate}, and energy demand~\citep{vanvliet2012vulnerability, auffhammer2017climate, fonseca2019seasonal, vanruijven2019amplification, decian2019global, deroubaix2021large, bloomfield2021quantifying}. Many of these studies report climate change impacts on the order of $\pm$ 5-10\% in long-term averages depending on the studied region, period, and part of the power system analysed~\citep{tobin2016climate, wohland2017more, karnauskas2018southward, tobin2018vulnerabilities, fonseca2019seasonal, decian2019global, pryor2020climate, bloomfield2021quantifying, deroubaix2021large}. Studies also report more pronounced impacts over shorter periods or smaller areas~\citep{tobin2016climate, tobin2018vulnerabilities, pryor2020climate, bloomfield2021quantifying, deroubaix2021large}. Although uncertainty surrounds climate change impacts on energy systems, ignoring climate change can be dangerous with respect to extremes and might lead to reduced system reliability~\citep{voisin2016vulnerability, harang2020incorporating, ralston2021climate, bennett2021extending} and resiliency. The potential consequences of these threats are underscored by recent real-world events, like reliability failures during rolling blackouts in California and Texas in 2020~\citep{CAISO2021rootcause} and 2021~\citep{mays2022private}, respectively, and resiliency failures during wildfires in the Western United States, Australia, and elsewhere~\citep{muhs2020wildfire,bill2022}. While peer-reviewed climate change attribution studies~\citep{stott2016climate, van2021pathways} have not been completed for these events, extreme weather driving these events (e.g., heat and drought) is projected to increase in severity and/or frequency under climate change in many parts of the world~\citep{coumou2012decade, aghakouchak2014global, zhuang2021quantifying, seneviratne2021weatherAR6}.

In recognition of these threats, a community of practice in energy-climate modelling has started to form that aims to better coordinate two types of models:
\begin{enumerate*}[label=(\arabic*)]
\item energy system models and
\item weather and climate models.
\end{enumerate*}
The community has formed around annual workshops~\citep{bloomfield2021importance,nextgenec20,nextgenec21}; monthly webinars~\citep{nextgenec22webinar}; overlapping research teams; and a virtual knowledge sharing platform enabling ongoing exchange about new research, code and data\footnote{The platform is hosted on Slack, see \url{https://join.slack.com/t/nextgenec/shared_invite/zt-13ylctgw5-OG7aWUAPQuC7dlRZkUszvA}.}. To expedite the development of this community, this perspective details the gaps between energy system and climate modellers, then suggests near- and long-term actions aimed at closing them. Our suggestions aim to bring the energy and climate modelling communities together to effectively and appropriately use climate information in guiding energy system design and operation.

\section{The Disconnect between Energy System and Climate Modelling Communities}
To highlight the gap between energy system and climate modelling communities, this section first describes the work of an illustrative member of each community. The description of the illustrative member is not meant to describe all members of each community, but is instead intended to provide a simplified representation of a typical member’s scope and priorities.

An illustrative energy system modeller uses one or more types of energy system models~\citep{pfenninger2014energy, ringkjob2018review} to better understand or advance the provision of reliable, affordable, and clean energy. Their energy system model is formulated to inform a particular decision, e.g. as an optimisation model to inform investment ~\citep{batlle2013electricity, schill2018long, rubino2021} or operational decisions ~\citep{rubino2021} under various objectives or as a simulation model to investigate system behaviour under certain boundary conditions~\citep{murphy2020resource}. Given the scale and complexity of real-world energy systems, their model sacrifices model formulation and spatio-temporal resolution to maintain computational tractability~\citep{nahmmacher2016carpe, pfenninger2017dealing, hilbers2019importance}. Oftentimes, this means formulating a deterministic problem and simplifying meteorological data across space or time, e.g. by ignoring climate-related uncertainty, modelling a few days per year, assuming perfect foresight, or aggregating sites or time periods~\citep{hoffmann2020review, kotzur2021modeler}. As deterministic models, many energy system models return the optimal planning decision or operational strategy for a given set of meteorological and other inputs.

\begin{figure}[t]
\centering
\includegraphics[width=0.9\textwidth]{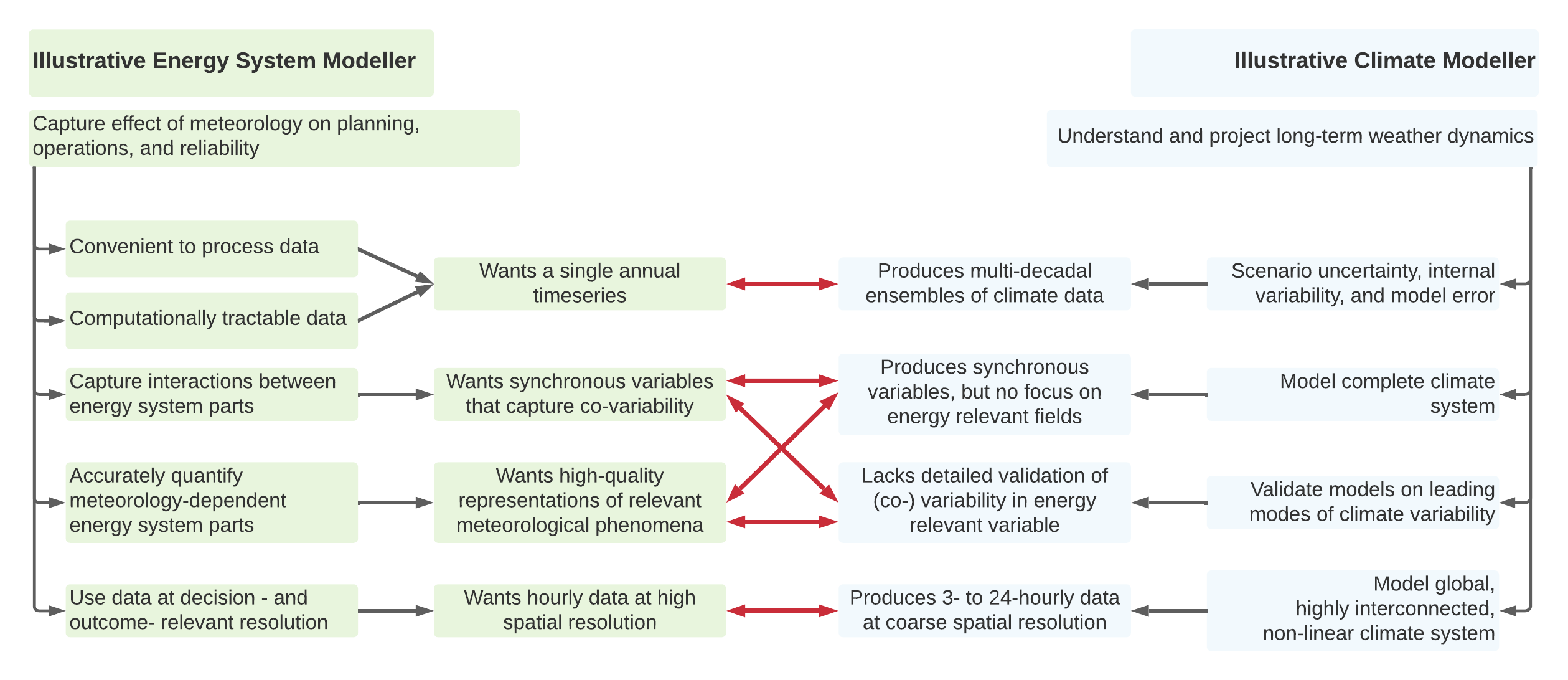}
\caption{Our illustrative energy system and climate modellers have different key priorities, leading to a disconnect (red arrows) between our illustrative energy system modeller needs and climate modeller outputs.}
\label{fig:fig1}
\end{figure}

In selecting meteorological data, our representative energy system modeller looks for five key attributes in a dataset:
\begin{enumerate*}[label=(\arabic*)]
\item the relevant spatio-temporal resolution for the studied decision;
\item synchronous with other climate-sensitive inputs, e.g. renewable production synchronous with electricity demand;
\item convenient to process;
\item computationally manageable; and
\item a high-quality representation of meteorological phenomena relevant to energy system and technology performance (see also Figure~\ref{fig:fig1}).
\end{enumerate*}
To satisfy these attributes, the status quo for meteorological data in energy system models is historical data, e.g. from reanalyses~\citep{hersbach2020era5, gelaro2017merra2, ramon2019comparison, urraca2018irradiance, kaspar2020reanalysis, bloomfield2021reanalysis} or reanalysis-derived products (e.g. Renewable.ninja~\citep{pfenninger2016pv, stafell2016wind}, EMHIRES~\citep{iratxe2016emhires, gonzalez2017emhires}, WIND Toolkit~\citep{draxl2015wind}, NSRDB~\citep{sengupta2018national} and the PECD~\citep{nuno2018timeseries,matteo2021pecd}). Historical meteorological datasets provide hourly (or even sub-hourly) data, which is the typical temporal resolution at which energy system models run when applied to planning and operational decisions ~\citep{deane2014impact}. Spatial resolutions in these datasets differ widely, but are generally on the order of 10 kms. With respect to synchronicity, many historical datasets include wind and solar resource data, which can be paired with historical electricity demand and hydropower time series. To process historical datasets, energy system modellers leverage widely-used, open-source code and/or tools (e.g., ECEM~\citep{troccoli2018creating}, Atlas~\citep{ruohomaki2018smart}, atlite~\citep{hofmann2021atlite}, and SAM~\citep{blair2018system}) that convert meteorological data, e.g. wind speeds, into energy system model inputs, e.g. wind power output. To maintain computational manageability, they use sub-decadal --- often even a single year --- of data. Finally, to understand the quality of the dataset, our illustrative energy system modeller might review papers validating and bias-correcting relevant variables from datasets against empirical data~\citep{pickering2020sub, cannon2015using, habte2017evaluation, king2014validation, bloomfield2021pattern}. They are, however, unlikely to extensively investigate the quality of the meteorological data for their specific purposes, nor to be fully aware of the in-depth meteorological literature assessing the representation of relevant meteorological processes and their vulnerability to climate change (e.g., \cite{woollings2010dynamical} for a European perspective and \cite{ravestein2018vulnerability}, \cite{gonzalez2019contribution} and \cite{pickering2020sub} for their connection to energy).

This selection process is much more daunting for meteorological datasets sourced from climate models seeking to represent the future (rather than historical) climate. To understand why, we first turn our attention to understanding the scope and priorities of an illustrative climate modeller.

Our illustrative climate modeller is primarily concerned with understanding and simulating long-term weather and climate change. They use general circulation models (GCMs), which employ a physics-based dynamical mathematical model of the circulation on Earth, to derive long-term climate and weather projections. They might further refine the accuracy and/or resolution of their weather projections by bias correcting and/or downscaling, which use statistical~\citep{abatzoglou2012comparison} or dynamical~\citep{losada2020potential} models to capture the effect of geography or other factors on weather that GCMs are too coarse to account for. To them, an energy system modeller is often seen as a downstream user of climate data: the climate modeller’s primary focus is to generate high-quality meteorological data, and to provide access to the raw meteorological output for downstream users. As a result, our climate modeller has little knowledge about data needs specific to the energy sector and does not necessarily assess the climate model’s skill in the context of energy system applications. Instead, in generating and publishing data, our climate modeller has several objectives. First, while being interested in how their model performs relative to observations and other models (e.g., \cite{eyring2019taking} and \cite{hausfather2020evaluating}), they likely care more about the model’s performance in purely meteorological-process terms  than about energy-relevant surface climate variables (e.g., they may focus on wind-speed at 10 m --- as at this level there are surface observations to validate models against --- rather than estimating wind-power capacity factors at turbine hub-height). Second, though some surface climate variables (or “Essential Climate Variables'') are extensively evaluated in climate models, that evaluation tends to focus on static statistical properties (e.g., mean, variance, and probability distributions) of meteorological variables in isolation rather than on time series of and co-variability between energy-system-relevant variables (e.g., in a \emph{kalte dunkelflaute}~\citep{van2019meteorological, li2021mesoscale}). Third, given the complex sources of uncertainty affecting climate models (e.g., scenario uncertainty, internal variability, and model error; \cite{hawkins2009potential}), our illustrative climate modeller typically views ensembles of many climate models as essential for understanding climate risk in present and future climate. Each simulation produced by them will therefore cover many decades including historical and future conditions. Ultimately, these datasets allow separation of fluctuations induced by climate change from natural climate fluctuations. Thus, overall our illustrative climate modeller might typically seek to draw conclusions from tens of model simulations (from one or more models), each of which will span 30 years or more for a historical and future climate (e.g., \cite{eyring2016overview}).

The expertise, scope, and priorities of our illustrative climate modeller significantly differ from those of our illustrative energy system modeller, leading to a disconnect in information and data flows between the two. As a result, when considering using outputs from our climate modeller, our energy system modeller faces a steep challenge in satisfying their five meteorological data priorities (see Figure~\ref{fig:fig1}). This steep challenge could --- and likely often does --- deter our energy system modeller from engaging with state-of-the-art climate datasets, thus leading to an over-reliance on historical meteorology despite its increasingly poor representation of the future.

With respect to resolution, our illustrative energy system modeller wants hourly data fields at high spatial resolution. Such high-resolution data is rarely if ever available, including in the ongoing CMIP6 High Resolution Model Intercomparison Project~\citep{haarsma2016}, which currently provides typically no more than 3-hourly data~\citep{eyring2016overview}. Our illustrative energy system modeller also lacks the knowledge, expertise, and/or resources to consider, select, and run appropriate downscaling and bias adjustment methods. Finding synchronous datasets is generally easier than finding datasets with appropriate resolution, as many climate datasets include all types of meteorological variables of interest for energy systems. However, available meteorological variables in climate datasets are often deficient, as discussed below. Additionally, projecting electricity demand and hydropower generation with bottom-up models and climate model outputs is an area of active research. Diverse methods and assumptions exist to model electricity demand, e.g. as a function of surface air temperature, electrification, and societal trends, indicating substantial methodological uncertainty~\citep{chang2021trends}. Providing accurate projections of hydropower generation potential given future meteorology requires complex hydrological modelling, but simpler approaches exist. Whether these simpler approaches are generalizable or remain valid in a changing climate is, however,  often unclear. For instance, profiles of hydropower generation potentials have been inferred using runoff in upstream basins~\citep{trondle2020trade,gotske2021future}, since generation potentials have been found to be highly correlated with reservoir inflow~\citep{liu2019high}. Finding convenient to process and computationally manageable datasets poses another significant challenge for the energy system modeller, to whom the concept of and processing tools for a meteorological ensemble are largely foreign. Standard climate change datasets of long-term ensembles radically differ from the single annual meteorological time series that our illustrative energy system modeller is trained to use. Finally, to understand whether a climate data set represents relevant meteorological phenomena well, an energy system modeller has two choices:
\begin{enumerate*}[label=(\arabic*)]
\item parse through literature, including documentation and journal articles, to understand the model’s skill with respect to each meteorological variable of interest and their co-variability, or
\item assess them themself.
\end{enumerate*}
The first choice is often not possible, as such information is typically unavailable and, where available, not quantified based on energy system model needs. But more importantly, our energy system modeller does not have the training or expertise to effectively carry forward either choice. Without familiarity with fundamental concepts in climate science, large parts of the existing literature are effectively useless.

Overall, these perspectives of an illustrative energy and climate modeller highlight several disconnects between energy and climate modelling that currently hinder effective usage of climate information in energy system modelling (see Figure~\ref{fig:fig1}). While some of these disconnects relate to data integration challenges (e.g., mismatches in spatio-temporal resolution), others relate to deeper challenges regarding the scope, objectives, expertise, and treatment of meteorological uncertainty embedded within each community.  In the next section, we offer near- and long-term actions that can help bridge these disconnects.

\section{Bridging the Divide through a New Approach to Data and Research}
The problems identified above currently prevent the use of the full potential of climate expertise and information in energy system modelling. To address these problems, we propose a set of near- and long-term interdisciplinary and transdisciplinary activities among the energy and climate modelling communities. In the near-term, our proposed interdisciplinary activities aim to expedite the use of future climate data in energy system modelling, generating much-needed insights for decision-makers. In the long term, our proposed transdisciplinary activities aim to enable two developments:
\begin{enumerate*}[label=(\roman*)]
\item energy-system-tailored climate datasets for historical and future meteorological conditions, and \item energy system models that can effectively leverage those datasets.
\end{enumerate*}
Proposed actions will require reframing and reconsidering methods and processes currently used to create and share data and knowledge between climate and energy modelling communities.

\subsection{Align Climate Model Outputs with Energy System Model Needs}
First, the climate modelling community should align the resolution of their outputs with energy system model needs. In future phases of coordinated climate simulations, climate modellers should output and save the following variables at hourly resolution to match the granularity of detailed energy system models: surface radiation (direct and diffuse); wind speeds and air density at multiple levels between 80 and 200m, including 100 m for intercomparison studies; surface air temperature; surface relative humidity; precipitation; runoff; and evaporation. With these variables, energy system modellers can estimate energy demand and potential electricity generation from wind, solar and hydropower with much greater fidelity (notwithstanding challenges associated with demand and hydropower projections discussed above). We acknowledge that for many current-generation global climate models, the benefits of producing hourly model output may be modest (compared to 3-hourly) but it is likely to increase in future models with higher resolution (as may already be the case for current regional climate models).  More importantly, access to hourly surface weather variables from climate models would greatly simplify the pathway to uptake of climate model data by energy-system specialists by
\begin{enumerate*}[label=(\roman*)]
\item improving comparability with historical reanalysis data like  ERA5;
\item easing integration with highly non-linear energy models (e.g., wind power has a cubic dependence on wind speed); and
\item directly matching the time-granularity of energy system models used in operations and planning.
\end{enumerate*}
Experience from previous projects, particularly the EU H2020 PRIMAVERA project, suggests that the additional storage costs of producing a small set of \emph{single level} surface weather data are modest compared to, e.g., archival of 3-D cloud physics datasets CFMIP or ISCCP~\citep{webb2017cloud} or Lagrangian storm track diagnostics~\citep{priestley2020overview}. Moreover, a greater availability of high frequency surface variables would benefit climate impact analyses in energy system modelling and other communities. Adopting hourly surface weather data output in the `standard' CMIP7 protocol would enable uptake in typical energy models and simplify comparison of system outcomes with historical versus future datasets, potentially enabling standardised energy risk assessments from CMIP7 onwards. As energy systems evolve, the suggested list of variables should be adopted, for instance, to reflect future changes in turbine hub height.

\subsection{Add Climate-Related Uncertainty Analysis to Default Energy System Modelling Toolbox}
Second, the energy system modelling community should add climate-related uncertainty quantification to their default toolbox. This requires accepting that there is not `one' representative meteorological time series they can use. The use of a single year of meteorological data, whether historical or future, is invalidated by the pronounced inter-annual variability in surface climate~\citep{bloomfield2016quantifying, wohland2018natural, lombardi2020policy}. Moreover, satellite-era reanalyses and other historical products are too short to capture multidecadal climate variability~\citep{wohland2019significant,wohland2021mitigating}. Single time series of one or more years carry the risk of missing low probability extreme events. The problem is substantially exacerbated in climate change assessment because multiple potential climate realisations from single- or multi-model ensembles must be considered, along with the role of model error. Diverse sources of uncertainty accumulate in these potential future climates, and the lack of future observation makes validation infeasible by definition.

Several potential strategies exist to include climate-related uncertainty in energy system models. Sensitivity analysis~\citep{ralston2021climate, schyska2021sensitivity}, advanced sampling procedures~\citep{nahmmacher2016carpe, pfenninger2017dealing, hilbers2019importance}, and robust decision-making~\citep{weaver2013improving, shortridge2016scenario, lempert2019robust} can capture climate-related uncertainty without modifying energy system model formulations. Advanced sampling aims to minimise the number of energy system model runs by reducing the ensemble size while still capturing its variability. Conversely, sensitivity analysis and robust decision-making would run energy system models for individual ensemble members to identify trade-offs between objectives, e.g. between cost and reliability. Other strategies like stochastic and robust optimization~\citep{kazemzadeh2019robust, perera2020quantifying, bennett2021extending} would embed climate uncertainty within the energy system model through major formulation changes. But these strategies face computational and parameterization challenges, constraining their applicability to real-world systems. With these enhanced uncertainty quantification strategies, the energy system community could begin to use ensembles of climate data~\citep{van_der_Wiel_2020}, so the energy system community should prioritise addressing remaining conceptual and computational challenges. Distributed modelling teams that incorporate experts in uncertainty quantification and climate science could partially alleviate the challenges with implementation of these uncertainty quantification strategies~\citep{decarolis2020leveraging}.

\subsection{Strengthen Transdisciplinary Collaborations between Energy System and Climate Modelling Communities}
\begin{figure}[ht]
\centering
\includegraphics[width=0.9\textwidth]{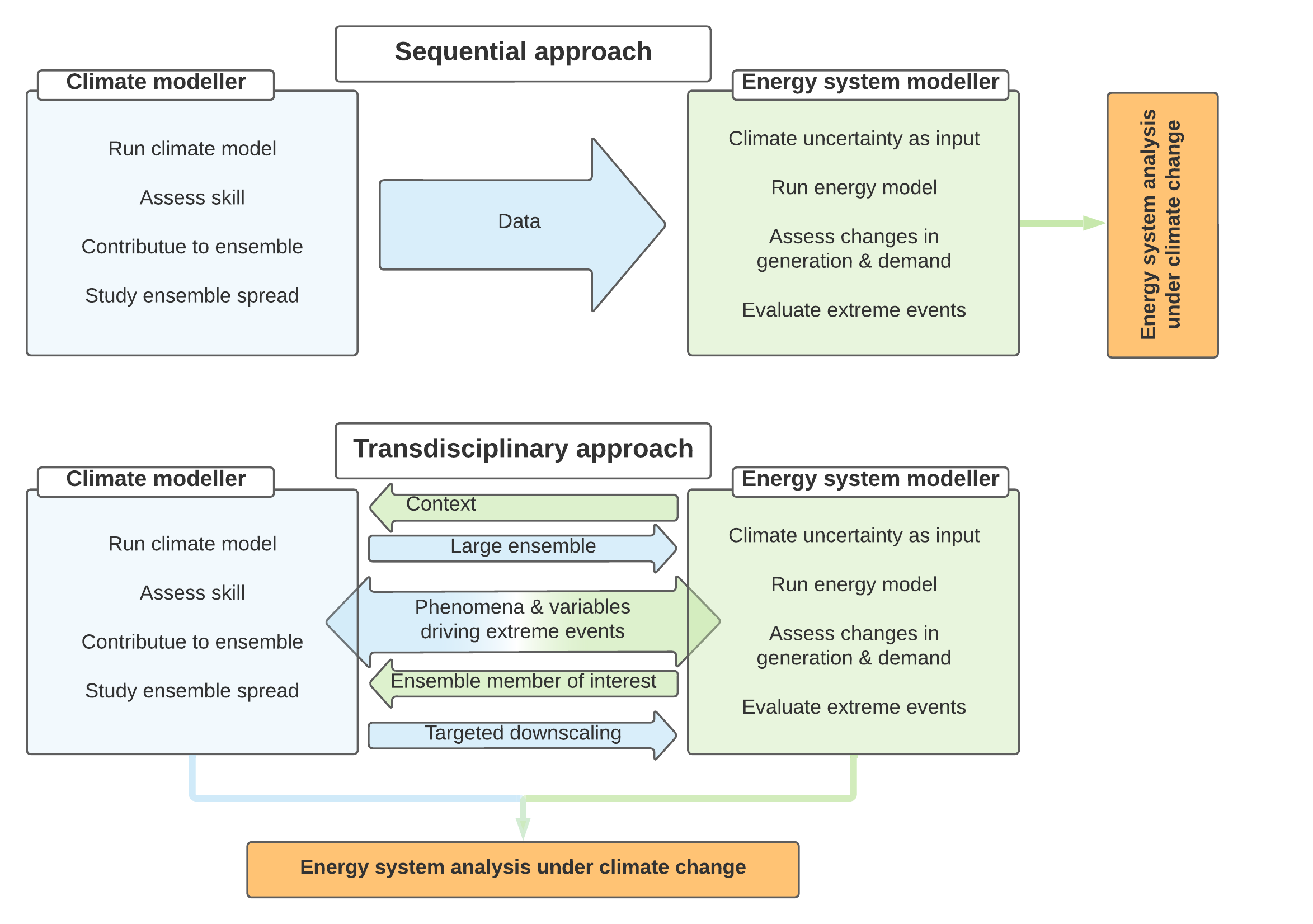}
\caption{Current sequential approach and the proposed transdisciplinary framework for improving coordination between energy system and climate modelling communities.}
\label{fig:fig2}
\end{figure}

Third, the energy system and climate modelling communities must engage more strongly in transdisciplinary collaboration (see Figure \ref{fig:fig2}). Using the full potential of climate information to guide climate change mitigation and adaptation will require coordination between the energy and climate modelling communities, taking into account uncertainties, limitations, and contexts. We acknowledge that climate services contribute to this coordination by seeking to provide useful climate information and context knowledge to impact modellers and policy makers through websites, interactive group activities and focused relationships (e.g., \cite{hewitt2017improving, fischer2021widening}). Although these services aim to overcome the disconnect between users and providers of climate information, substantial issues persist. For instance, experts report that climate services mainly focus on delivering better data rather than advancing the needed research identified in this perspective~\citep{findlater2021climate}.  In short, there is a need for the two communities to work together rather than merely seeking an efficient interface across which pre-packaged information is passed.

Sequential approaches that divide labour by expertise are problematic because they assume that the climate and energy steps can be separated. For example, when investigating the impacts of extreme events on an energy system, a climate modeller could try to identify the most and least extreme ensemble members which are subsequently used by the energy expert in their modelling. This approach, however, only works if the climate modeller's expertise is sufficient to predict what the energy model will consider extreme~\citep{van_der_Wiel_2020}. This is unlikely given the complexity of energy models with hundreds of millions of decision variables and interactions between technologies and sectors.

Instead of a sequential approach, we propose an iterative transdisciplinary framework (see Figure \ref{fig:fig2}). In this framework, climate modellers would generate climate ensembles. Climate and energy modellers would then jointly evaluate those ensemble members to identify which ensembles, variables, locations, and meteorological phenomena yield extreme energy system outcomes (e.g., non-served or peak residual demand; \cite{van2019meteorological}). This identification would occur by running each ensemble member through energy system models~\citep{van_der_Wiel_2020}. Identified variables and phenomena that yield extreme power system outcomes, as opposed to just extreme meteorological conditions, would be fed back to the climate modelling community to enable better skill assessment and targeted model development for energy system applications. Identified variables and phenomena would also guide downscaling activities to produce energy system inputs, which would be fed back into energy system model development.

Research funders can aid this transdisciplinary collaboration if the limitations of traditional paradigms are reflected in their priorities and objectives. Emphasis on enhanced interdisciplinary training and exchange, alongside career support for early- and mid-career researchers, would support those whose research already spans this interdisciplinary space. Moreover, funding agencies should push transdisciplinary research in energy and climate science, particularly where such research challenges the prevailing paradigms. To appropriately assess transdisciplinary  proposals, funders should actively  mitigate the risk that reviewers approach transdisciplinary work from the perspective of a single discipline by implementing policies that embed transdisciplinary diversity and awareness into the entire review process. Funding opportunities in this area should support fundamental scientific challenges of translating climate risk into energy systems (e.g., maintaining computational tractability and understanding the roles of, as well as mitigating against, different sources of uncertainty) alongside larger-scale ‘applied’ or ‘solutions-oriented’ programmes

\section{Conclusion}
The energy system modelling community faces a transformation: accounting for the non-stationary meteorological conditions associated with climate change. Succeeding in this transformation is crucial for future energy systems to provide reliable, affordable, and clean energy under an uncertain future climate. Yet, this transformation is held back by a disconnect between energy and climate modelling communities.

To bridge this disconnect and to overcome the challenges associated with modelling energy systems under climate change, this article explored the underlying drivers, then proposed three inter- and transdisciplinary actions:
\begin{enumerate*}[label=(\arabic*)]
\item the climate modelling community should align output variables and their spatio-temporal resolutions with energy system modelling needs;
\item the energy system modelling community should add climate-related uncertainty quantification to their default analytical toolbox; and
\item both communities should engage in a transdisciplinary approach that ensures the development and evaluation of climate information in line with energy sector needs.
\end{enumerate*}
Ultimately, these actions point to the fact that effective energy system modelling in a changing climate should not continue to treat climate and energy modelling sequentially. Rather, effective energy system modelling must understand climate and energy modelling as a transdisciplinary interactive endeavour. Given intensifying climate change and rapid decarbonization of energy systems, the time for this endeavour is now.

\backmatter

\bmhead{Acknowledgments}
The hosting of the NextGenEC workshops – which led to the initiation and progression of this paper – was enabled with the kind support of the University of Reading and the National Centre for Atmospheric Science (e.g., through access to conferencing software), and the intitial workshop concept was developed during the PRIMAVERA project (which received funding from the European Union's Horizon 2020 Research and Innovation Programme under grant agreement no. 641727).

\bmhead{Funding}
The authors would like to acknowledge the following sources of funding.
During large parts of this work, Jan Wohland was funded through an ETH Postdoctoral Fellowship and acknowledged support from the ETH and Uniscientia foundations.
Laurens P. Stoop received funding from the Netherlands Organisation for Scientific Research (NWO) under grant number 647.003.005 and is part of the IS-ENES3 project that has received funding from the European Union’s Horizon 2020 research and innovation programme under grant agreement No. 824084.
Bryn Pickering was funded by the Swiss Federal Office for Energy (SFOE) under grant number SI/502229.
Chris Dent is part of the projects "Managing Uncertainty in Government Modelling" (sponsored by the Alan Turing Institute) and "Decision support under climate uncertainty for energy security and net zero " (sponsored by the Alan Turing Institute and EPSRC).
Alexander Grochowicz acknowledges UiO:Energi Thematic Research Group Spatial-Temporal Uncertainty in Energy Systems (SPATUS).
Katharina Gruber is funded through the reFUEL project, a ERC-grant with No. ERC2017-STG 758149.
Philip H\"artel is part of the PROGRESS project funded by the Federal Ministry for Economic Affairs and Energy (BMWi.IIC5, funding reference 03EI1027).
Julie K. Lundquist has funding provided by the US Department of Energy Office of Energy Efficiency and Renewable Energy Wind Energy Technologies Office.  Copyright statement: This work was authored in part by the National Renewable Energy Laboratory, operated by Alliance for Sustainable Energy, LLC, for the U.S. Department of Energy (DOE) under Contract No. DE-AC36-08GO28308. The U.S. Government retains and the publisher, by accepting the article for publication, acknowledges that the U.S. Government retains a nonexclusive, paid-up, irrevocable, worldwide license to publish or reproduce the published form of this work, or allow others to do so, for U.S. Government purposes.
David J. Brayshaw was a Co-Principal Investigator on the EU H2020 PRIMAVERA project (grant number 641727) and is the lead-convenor of the Next Generation Challenges in Energy Climate Modelling workshops from which this perspective was developed.

\bmhead{Authors' contributions}
Conceptualization: \emph{David J. Brayshaw, Jan Wohland, Michael T. Craig, and Laurens P. Stoop}, Original draft: \emph{Jan Wohland, Michael T. Craig, and Laurens P. Stoop}, Revision, review, and editing core group: \emph{David J. Brayshaw, Michael T. Craig, Alex Kies, Bryn Pickering, Laurens P. Stoop and Jan Wohland}, Revision and review: \emph{All authors}. All authors (except L. Kotzur) were present at the Next Generation Energy Climate Modelling 2021 workshop that led to the conceptualization of this paper and was organized by David J. Brayshaw. The ordering of the first two authors was decided by a coin toss; these co-first authors can prioritise their names as first authors when adding this paper’s reference to their r\'esum\'es.

\bibliography{ms.bib}

\end{document}